\documentclass[aip,jcp,reprint]{revtex4-1}

\usepackage{amsfonts, amsmath, amssymb,latexsym}
\usepackage{subfigure}
\usepackage{graphicx}

\newcommand{\me}{\mathrm{e}}

\newcommand{\dif}{\mathrm{d}}

\begin{document}

\title{Pairing of 1-hexyl-3-methylimidazolium and tetrafluoroborate ions in  $n$-pentanol}

\author{P. Zhu}
\email{zpeixi@tulane.edu}

\author{L. R. Pratt} 
\email{lpratt@tulane.edu}
\author{K. D. Papadopoulos}
\email{kyriakos@tulane.edu}
\affiliation{Department of Chemical and Biomolecular Engineering, Tulane
University, New Orleans, LA 70118}

\date{\today}

\begin{abstract} 
Molecular dynamics simulations are obtained and analyzed to study
pairing of 1-hexyl-3-methylimidazolium and tetrafluoroborate ions in 
$n$-pentanol, in particular by evaluating the potential-of-mean-force
between counter ions. The present molecular model and simulation
accurately predicts  the dissociation constant $K_{\mathrm{d}}$ in
comparison to experiment, and thus the behavior and magnitudes for the
ion-pair pmf at molecular distances, even though the dielectric constant of the
simulated solvent differs from the experimental value by about 30\%.
A naive dielectric model does not capture molecule structural effects such
as multiple conformations and binding geometries of the Hmim$^+$ and
BF$_4{}^-$ ion-pairs.  Mobilities identify
multiple time-scale effects in the autocorrelation of the random forces
on the ions, and specifically a slow, exponential time-decay of those
long-ranged forces associated here with dielectric friction effects.
\end{abstract}

\maketitle 

\section{Introduction} 
Ion-pair encounter, binding, and dissociation is central to solution
chemistry, including chemistry in organic solvents,\cite{Welton:1999ul} and to
understanding specific electrolyte solutions in a wide range of
practical settings.\cite{Kunz:2010}  We recently obtained an
experimental determination of ion-pairing specifically of
1-hexyl-3-methylimidazolium tetrafluoroborate in $n$-pentanol.\cite{Zhu:2009cm}
Those results encouraged us to pursue detailed  testing of the
molecular-scale description  of ion pairing and dynamics for that
system.  We report results of that testing here.   We report, among other
results, evaluation of memory functions for ion mobility that 
provide a direct signature of long-ranged ion-solution
interactions,\cite{Wolynes:1978kc} a signature that we had not
anticipated in advance of these simulations.

The theoretical testing naturally relies heavily on molecular
simulations that have become accessible in recent years. [Details of the
present simulation calculations are provided in the Appendix.]  The
targets of these calculations were the ion-pair potential-of-mean-force
(pmf), the ion-pairing dissociation constant implied by that pmf, then
further the mean-square-displacement of the ions individually, their
velocity autocorrelation functions, and the corresponding memory
functions.  To assist the interpretation of these quantities, we also
evaluated the dielectric constant of the solvent in the present
simulation model.

\begin{figure}
         \begin{center}
             \includegraphics[width=3.0in]{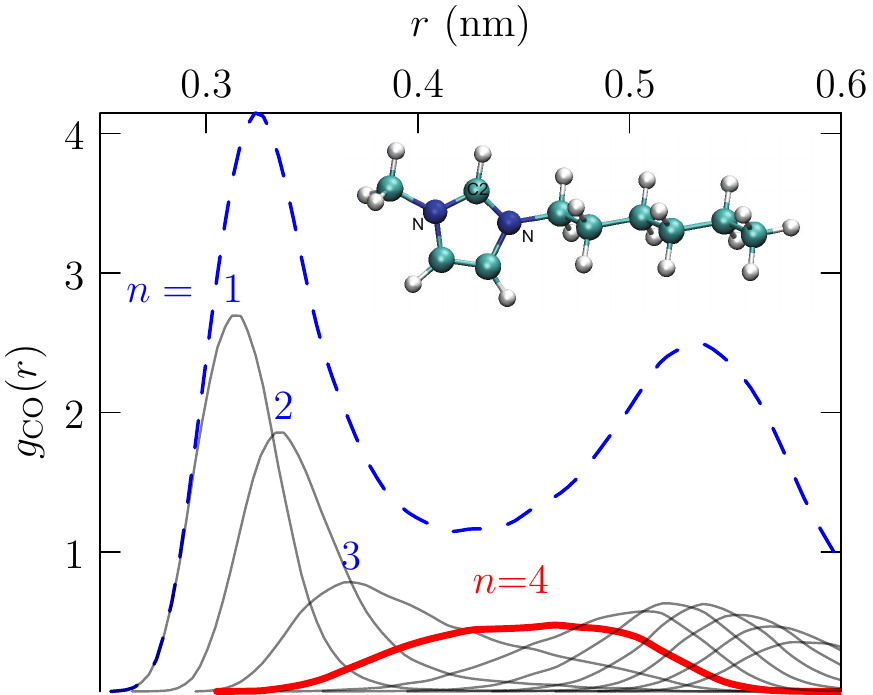}
             \includegraphics[width=3.0in]{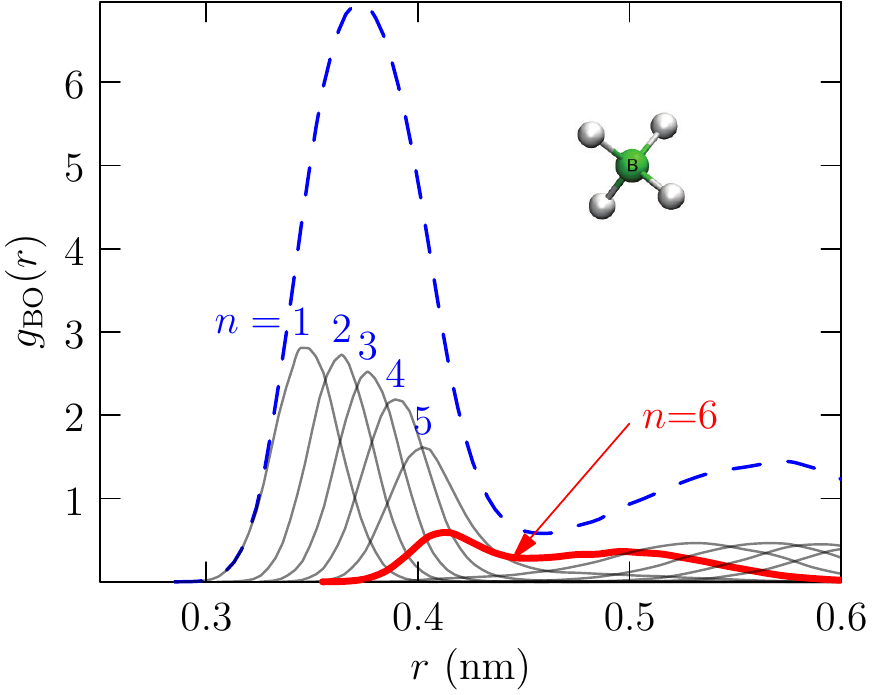}
         \end{center}
         \caption{Radial distribution functions of $n$-pentanol O-atom from the ion 
         centers (upper) middle C-atom of the imidazolium ring and (lower) the B-atom 
         of BF$_4{}^-$, together with the distance-ordered decompositions (grey).  For the upper
         panel $n$ = 1, 2, 3  fills out the principal peak and $n$=4 participates in both
         first and second shell.  For the lower panel $n$ = 1, \ldots , 5 fills out the principal peak and $n$=6 participates in both
         first and second shell.
         \label{fig:IL_structure}
         }
\end{figure}

\section{Results and Discussion}
The analysis requires that we choose a center for the molecular ions of interest.
Radial distribution functions associated with the chosen centers,
(Fig.~\ref{fig:IL_structure}) mid-C for the 1-hexyl-3-methylimidazolium
(Hmim$^+$) ion and B for the tetrafluoroborate (BF$_4{}^-$) ion, 
characterize ion solvation.

\begin{figure}
         \begin{center}
             \includegraphics[width=3.0in]{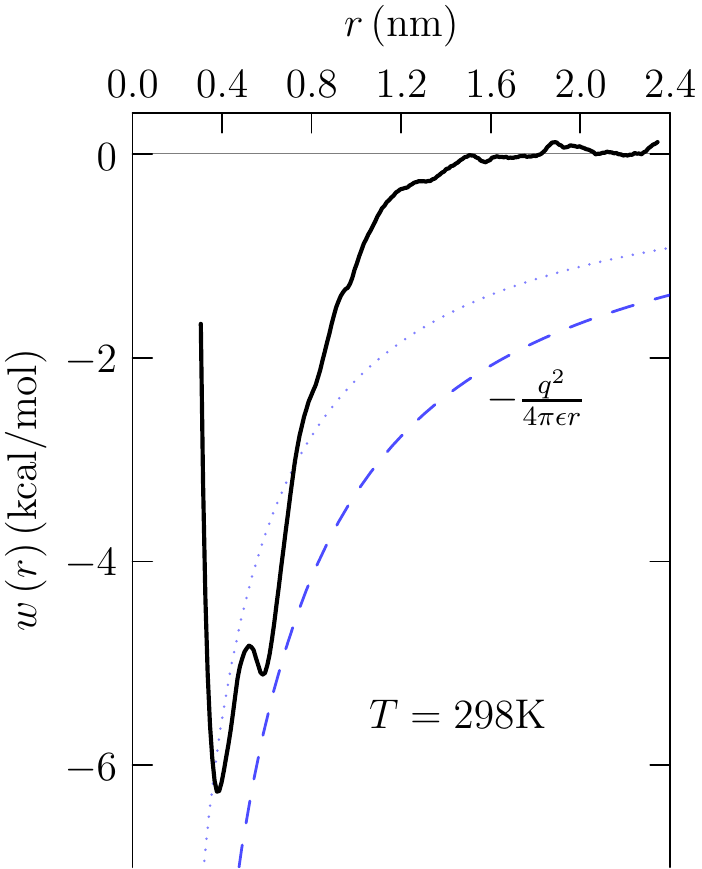}
             \includegraphics[width=3.0in]{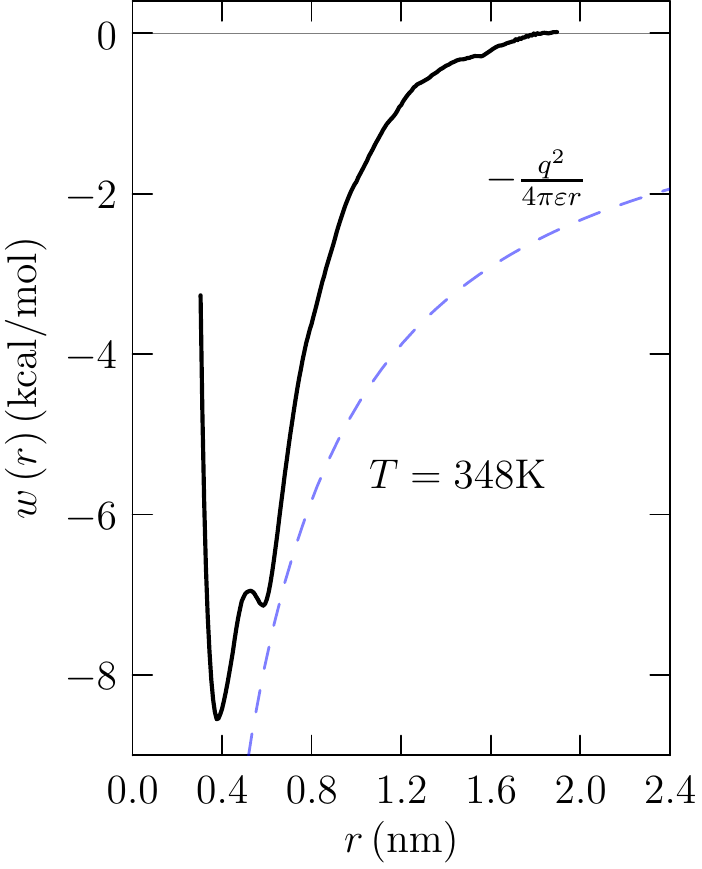}
         \end{center}
         \caption{The CB (Fig.~\ref{fig:IL_structure}) pmf in $n$-pentanol as a function
          of the CB radial displacement.   For the modelled $n$-pentanol solvent, the static
         dielectric constants are $\epsilon/\epsilon_0$ = 7.1 and 10.0
         at $p$ = 1~atm and $T$ = 298.15~K and 348.15~K, respectively.  The dotted
         curve utilizes the experimental dielectric constant (15).\cite{CRC}
         \label{fig:pmf}}
\end{figure}

The CB-pair pmf, $w(r),$ (Fig.~\ref{fig:pmf}) shows strong association.
Comparison with a naive dielectric continuum model, $w(r) \approx -
q^2/4\pi\epsilon r,$ highlights the molecular structure of the
simulation result.   The local minimum for $r \approx 0.65$~nm 
reflects the binding of the BF$_4{}^-$ ion to the ring \emph{center} of
the Hmim$^+$ ion.  Thus multiple binding geometries can be a specific
complication in pairing of molecular ions.  Additionally, the
ionic electric charge is significantly distributed
over these molecular structures, so this dielectric model (Fig.~\ref{fig:pmf})
is indeed naive for that reason also. 

\begin{figure}
         \begin{center}
             \includegraphics[width=3.0in]{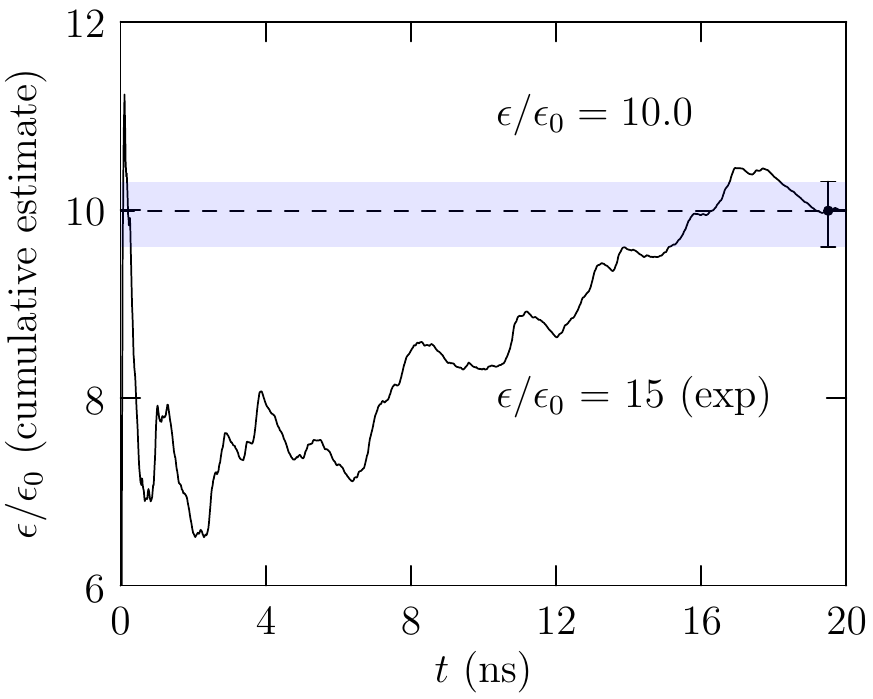}
         \end{center}
         \caption{Computed dielectric constant for the simulated
         $n$-pentanol liquid at $T$ = 298.15K and $p$=1~atm.  The shaded
         band indicates the 95\% confidence interval [9.6,10.3],
         established by bootstrap resampling. See Sec.~\ref{sec:dielectric}.
         \label{fig:dielectric}}
\end{figure}

At the longest range here, the computed pmf is foreshortened due to the
periodic boundary conditions, \emph{i.e.,} the mean forces between the
ions along a cartesian axis will be zero at a boundary face for the
simulated result but not for the dielectric model, which is expected to
be correct at the longest range. Recognizing that distinction, the
dielectric model utilizing the dielectric constant evaluated for the
simulation model $n$-pentanol solvent (Fig.~\ref{fig:dielectric})
over-estimates the maximum binding free energy of this system.
This naive dielectric model predicts a more accurate
maximum binding free energy when the experimental dielectric constant is
utilized, but that comparison has no physical significance.

\begin{figure}
         \begin{center}
             \includegraphics[width=3.0in]{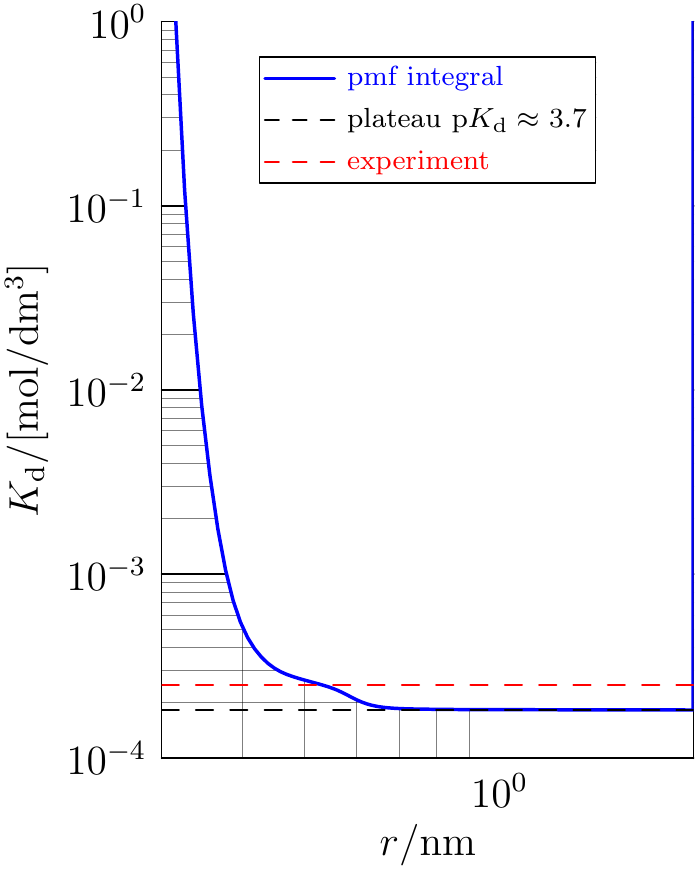}
         \end{center}
         \caption{Evaluation of the dissociation constant
         $K_{\mathrm{d}}$ of Eq.~\eqref{eq:Kd} and comparison with experiment.\cite{Zhu:2009cm}  The defining integral
         operationally plateaus for $r>0.65$~nm, and closely agrees with
         the experimental result.   The ripple near
         $r \approx 0.6$~nm reflects the second (outer or ring binding) geometry
         identified with Fig.~\ref{fig:pmf}, and thus that second
         binding mode affects the predicted $K_{\mathrm{d}}$.
         \label{fig:Kd02}}
\end{figure}

The dissociation equilibrium ratio  is defined by reference
to the equilibrium
\begin{eqnarray}
\mathrm{Hmim \cdot BF_4} \leftrightharpoons \mathrm{Hmim}^+ + \mathrm{BF_4}^-~,
\end{eqnarray}
and then
\begin{eqnarray}
K_{\mathrm{d}}  = \frac{\left(\rho_{\mathrm{Hmim^+}}\right)\left(\rho_{\mathrm{BF_4{}^-}}\right)}{\rho_{\mathrm{Hmim \cdot BF_4}}}
\label{eq:Kdratio}
\end{eqnarray}
with $\rho_\alpha$ the number density of species $\alpha$.  The species
indicated in Eq.~\eqref{eq:Kdratio} are identified by determining
whether the mid-C \ldots B atom pairs are within a radius $r$ (paired)
or not (unpaired). Basic statistical thermodynamics  identifies this
ratio\cite{BPP,Chempath:2009ws,Asthagiri:2010tj} as
\begin{eqnarray}
\frac{1}{K_{\mathrm{d}}} = \frac{p(n=1)}{p(n=0)\rho_{\mathrm{BF_4{}^-}} }~,
\label{eq:BPP}
\end{eqnarray}
and we acknowledge that $\rho_{\mathrm{BF_4{}^-}}  = \rho_{\mathrm{Hmim^+}}$. 
$p(n)$ is the probability of observing $n$
mid-C atoms of Hmim$^+$ ions within $r$ of the B-atom of a distinguished
BF$_4{}^-$ ion.   For the infinite dilution circumstances of the present
study, the ratio of probabilities (Eq.~\eqref{eq:BPP}) is given precisely
by the Poisson distribution formula\cite{Zhu:2011wn}
\begin{eqnarray}
\frac{p(n=1)}{p(n=0)\rho_{\mathrm{BF_4{}^-}} } =  
4 \pi \int_0^r \exp\left\lbrack-w(r^\prime)/k_\mathrm{B}T\right\rbrack r^\prime{} ^2 \dif r^\prime ~.
\label{eq:Kd}
\end{eqnarray}
If the lengths on the right of Eq.~\eqref{eq:Kd} are in nm, and if
$K_{\mathrm{d}}$ is mol/dm$^3$, the  factor for conversion of units is
6.023$\times 10^{23}/10^{24}$ = 0.6023 dm$^3$/(mol nm$^3$), so
\begin{eqnarray}
K_{\mathrm{d}}= \frac{1}{(0.6023)\times
4 \pi \int_0^r \exp\left\lbrack-w(r^\prime)/k_\mathrm{B}T\right\rbrack r^\prime{} ^2 \dif r^\prime} ~.
\label{eq:Kdnext}
\end{eqnarray}
The predicted dissociation constant
(Fig.~\ref{fig:Kd02}) operationally plateaus for $r>0.65$~nm, and
closely agrees with the experimental result.   The ripple near $r\approx
0.6$~nm reflects the second (outer or ring binding) geometry identified
with Fig.~\ref{fig:pmf}, and thus that second binding mode affects the
predicted $K_{\mathrm{d}}$. Overall, we  conclude that the present
molecular model and simulation predicts correct behavior and magnitudes
for the pmf at molecular distances, even though that pmf is
foreshortened at long-range as noted above.

\begin{figure}
         \begin{center}
             \includegraphics[width=3.0in]{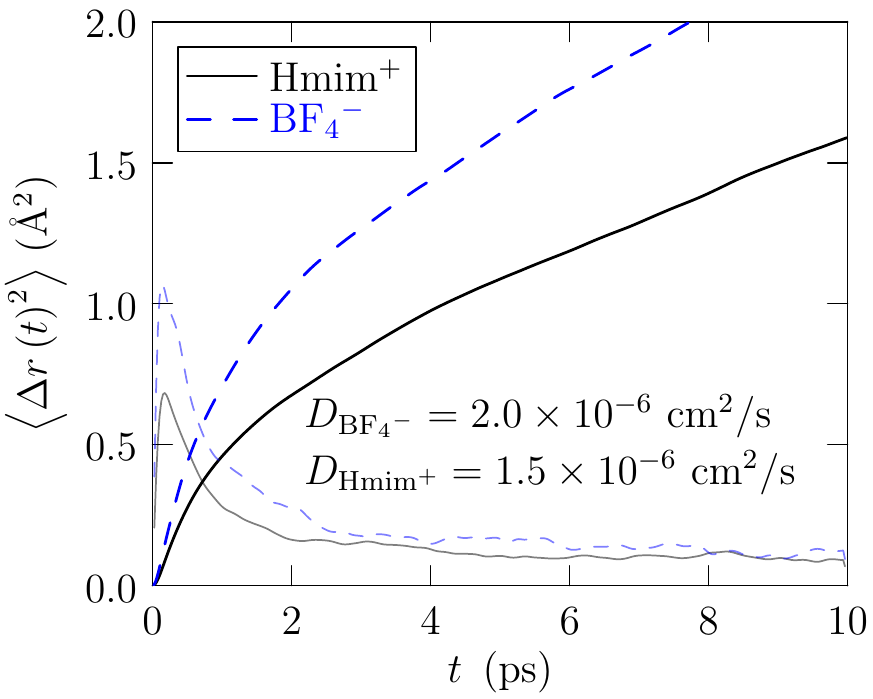}
         \end{center}
         \caption{Mean-square-displacements of the center-of-mass of
         these molecular ions observed in simulations of the individual
         ions in $n$-pentanol.  The fainter peaked curves in the
         background are the time-derivatives of the
         mean-square-displacements obtained from Eq.~\eqref{eq:msd}.
         The experimental value\cite{Zhu:2009cm} for 
         the average diffusivity is 1.87$\times 10^{-6}$ cm$^2$/s.
         \label{fig:MSD01}}
\end{figure}

\begin{figure}
         \begin{center}
             \includegraphics[width=3.0in]{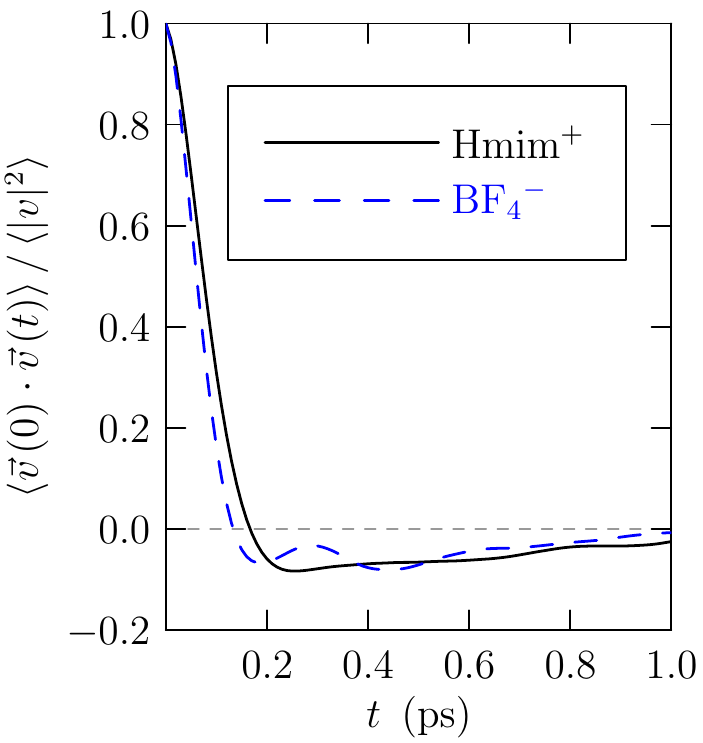}
         \end{center}
         \caption{Autocorrelation functions of the center-of-mass velocities of
         these molecular ions.  The time interval shown here is not sufficient
         to obtain the self-diffusion coefficients of Fig. ~\ref{fig:MSD01}.
         On this time interval these functions are qualitatively similar
         even though  the molecular structures of these ions are qualitatively
         different.
         \label{fig:vacf}}
\end{figure}

The kinetics associated with the mobilities of these ions were
characterized first on the basis of the observed
mean-square-displacements (Fig.~\ref{fig:MSD01}) 
\begin{eqnarray}
\frac{\dif \left\langle \Delta
r\left(t\right)^2\right\rangle}{\dif t} = 2 \int_0^t \left\langle \vec{v}\left(0\right)\cdot \vec{v}\left(\tau\right)\right\rangle \dif\tau~,
\label{eq:msd} \end{eqnarray}  
with the underlying velocity
autocorrelation functions (Fig.~\ref{fig:vacf})
\begin{eqnarray}
C(t) = \left\langle \vec{v}\left(0\right)\cdot \vec{v}\left(t\right)\right\rangle /\left\langle \vert v\vert^2\right\rangle~.
\end{eqnarray}
The indicated velocities are those of the center-of-mass of the extended
molecular ions.  We also considered a memory function $\gamma(t)$ defined by\cite{RWZ2001}
\begin{eqnarray}
m \frac{\dif C(t)}{\dif t}  = -\int_0^t \gamma(t-\tau)C(\tau) \dif \tau~,
\label{eq:memoryfunction}
\end{eqnarray}
with $m$ the mass of the ion. $\gamma(t)$ provides the autocorrelation
of the random forces on the ions.  We emphasize the connection to the forces 
with the notation $\Omega^2 = \left\langle F^2\right
\rangle/3mk_{\mathrm{B}}T$ so that
\begin{eqnarray}
\gamma(0) = m\Omega^2~.
\end{eqnarray}We extracted $\gamma(t)$ from $C(t)$
on the basis of the relation \cite{RWZ2001}
\begin{eqnarray}
\int_0^\infty \me^{-st}\gamma(t)dt 
\equiv \tilde{\gamma}(s) =     m\left(\frac{1}{\tilde C(s)} - s\right)~,
\end{eqnarray}
which follows from Eq.~\eqref{eq:memoryfunction}, with ${\tilde C(s)} $ the Laplace transform of $C(t)$. 

\begin{figure}
         \begin{center}
             \includegraphics[width=3.0in]{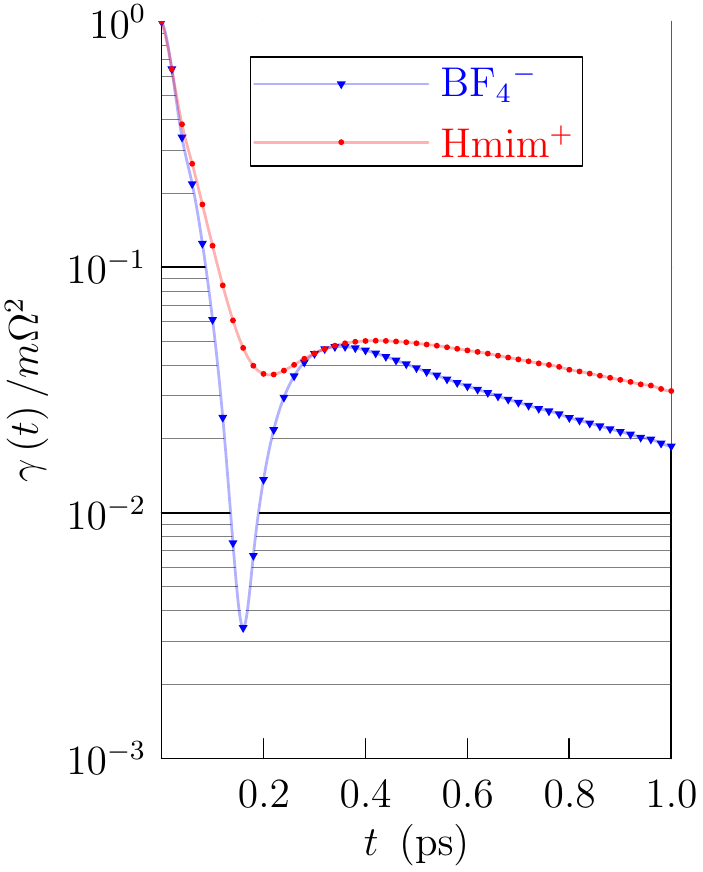}
         \end{center}
         \caption{Logarithm of the normalized autocorrelation function
         of the random forces on these ions, obtained with the Stehfest
         algorithm\cite{Stehfest:1970vj} from the velocity
         autocorrelation functions (Eq.~\eqref{fig:vacf}).   That
         $\gamma(t)$ is non-negative here is obvious but special.
         $\gamma(t)$ decays approximately exponentially for the largest
         times shown.   This long-time behavior was suggested many years
         ago in considering dielectric friction on ion
         mobilities.\cite{Wolynes:1978kc}   This behavior seems not to
         be have been observed until now, but Annapureddy and Dang have
         also obtained the analogous result for the autocorrelation of
         the forces on stationary alkali metal ions in
         water.\cite{Annapureddy}
         \label{fig:LogGamma}}
\end{figure}

We utilized the Stehfest  algorithm\cite{Stehfest:1970vj}
to invert the Laplace transform numerically.
$\gamma(t)$ is found  (Fig.~\ref{fig:LogGamma}) to be non-negative
in these cases.  Furthermore, $\gamma(t)$ displays two different time
regimes, and specifically a slow, exponential time-decay for the longest
times analyzed here.   The long-time decay recalls the discussion of
Wolynes many years ago\cite{Wolynes:1978kc}  of the effects of
dielectric friction on ion mobilities.    This behavior seems not to be
have been observed until now, but Annapureddy and Dang have
obtained the analogous result for the autocorrelation of the forces on
stationary alkali metal ions in water.\cite{Annapureddy}

\section{Conclusion}

The present molecular model and simulation accurately predicts  the
dissociation constant $K_{\mathrm{d}}$ of 1-hexyl-3-methylimidazolium
and tetrafluoroborate ions in  $n$-pentanol in comparison to experiment,
and thus the behavior and magnitudes for the ion-pair pmf at molecular
distances, even though the dielectric constant of the simulated solvent differs
from the experimental value by about 30\%. A naive dielectric model does
not capture molecule structural effects such as multiple conformations
and binding geometries of the Hmim$^+$ and BF$_4{}^-$ ion-pairs. 
Mobilities identify multiple time-scale
effects in the autocorrelation of the random forces on the ions, and
specifically the slow, exponential time-decay of those long-ranged forces
associated here with dielectric friction effects.

\section{Acknowledgements}
This work was supported by the National Science Foundation under the NSF
EPSCoR Cooperative Agreement No. EPS-1003897, with additional support
from the Louisiana Board of Regents.  Support from the Louisiana Board
of Regents Industrial Ties Research Subprogram (ITRS) and Chevron Energy Technology Company is gratefully acknowledged.

\section{Appendix:  Methods}

Potential parameters of ionic liquid (IL) 1-hexyl-3-methylimidazolium
tetrafluoroborate were taken from de Andrade,\cite{deAndrade:2002de}
which is the AMBER force field with slight modification. $n$-Pentanol force
field parameters were taken from AMBER.\cite{Hornak:2006gx} Partial atom
charges of Hmim$^+$ were derived following de Andrade's procedure, while
those of BF$_4{}^-$  were taken directly from their
work.\cite{deAndrade:2002de}  Partial atom charges used in this work for
$n$-pentanol were those developed by Kuhn.\cite{Kuhn:2002ed}

Our systems were simulated using  the AMBER~10 package in an
isothermal-isobaric ensemble (NPT) with periodic boundary conditions.
The cutoff for nonbonded interactions was 1.7~nm,  and Ewald summation
was used to calculate electrostatic interactions. The temperature was
regulated with Langevin dynamics, while pressure was controlled by
Berendsen's weak coupling algorithm.  All C--H bonds were constrained by
SHAKE algorithm.\cite{RYCKAERT:1977gp} For the simulation of IL pair in
$n$-pentanol, an ion pair was first equilibrated without cutoff in
vacuum for 10~ns to get an optimized geometry. Then it was placed in the
center of a (4.8~nm)$^3$ cubic box, which was packed
uniformly with 620 $n$-pentanol molecules using
Packmol.\cite{Martinez:2009di} For the single ion/$n$-pentanol systems,
Hmim$^+$ or BF$_4{}^-$ was simply wrapped up with 620 $n$-pentanol
molecules in a cubic box of the same size. Aging was 2~ns  at 298.15~K
under 1~atm with 1~fs integration time step. Then 8~ns production
equilibrium run was performed. For the single ion/$n$-pentanol systems,
the production runs extended to 10~ns. Configurations were saved every 1~ps
for further analysis. For the velocity autocorrelation function and the
mean square displacement of each ion,  1~ns
trajectories, with a time step of 2~fs, were obtained, saving the phase
point at each 10~fs.

\subsection{The pmf and WHAM calculations}
To obtain the ion-pair pmf, a reaction coordinate was defined as
the distance between the central carbon of the imidazolium ring
(Fig.~\ref{fig:IL_structure}) and boron atom of the BF$_4{}^-$ ion.
Window calculations then utilized a harmonic stratifying potential with
a force constant of 100~kcal/(mol$\cdot$nm) for windows from 0.3~nm to 2.35~nm (at
298.15K), and from 0.3~nm to 1.95~nm (at 348.15K), with an increment of
0.05~nm. Typically, the system was equilibrated for 1~ns to initiate
each window simulation, and then followed by 10~ns
production run.  For window separations greater than 1.7~nm,
the initial configuration for the next window was taken
from the last configuration from the previous MD simulation.  The weighted histogram analysis method
(WHAM)\cite{Grossfield:WHAM} was used to synthesize the final pmf profile.
For the calculation at 348.15~K,  the system consisted of an IL
pair and 340 $n$-pentanol molecules in a (4~nm)$^3$ cubic
cell.

\subsection{$n$-Pentanol Dielectric Constant}\label{sec:dielectric}
Assessment of the static dielectric constant of liquid $n$-pentanol is
required in considering the molecular pair potential of the mean forces
for Hmim$^+$ \ldots BF$_4{}^-$ in $n$-pentanol. The dielectric constant
for this model was obtained by standard simulation
methods.\cite{Wu:2006bg,Yang:2010hd} The calculation treated 620
$n$-pentanol molecules under standard periodic boundary conditions. The
results ($\epsilon/\epsilon_0$ = 7.1 and 10.0) were extracted from
averaging over 20~ns of simulation trajectory under constant pressure
conditions, and at constant temperatures of 298.15~K and 348.15~K,
respectively.   These computed values for the dielectric constant of
the modelled solvent are about 30\% smaller than experimental
values (Fig.~\ref{fig:dielectric}).

\clearpage



%

\end{document}